\documentclass[preprint,superscriptaddress,nofootinbib]{revtex4}
\usepackage[dvips]{graphicx}
\usepackage[usenames]{color}
\usepackage{setspace}
\usepackage{amsmath}
\usepackage{amssymb}
\usepackage{amsthm}
\usepackage{epsf}
\usepackage{epsfig}
\usepackage{hyperref}

\begin{document}


\preprint{UT-HET-118}
\preprint{EPHOU-16-017}

\title{
  Phenomenological signatures of mixed complex scalar WIMP dark matter
}
\author{Mitsuru~Kakizaki}
\email{kakizaki@sci.u-toyama.ac.jp}
\affiliation{
  Department of Physics, University of Toyama, 
  3190 Gofuku, Toyama 930-8555, Japan
}
\author{Akiteru~Santa} 
\email{santa@jodo.sci.u-toyama.ac.jp}
\affiliation{
  Department of Physics, University of Toyama, 
  3190 Gofuku, Toyama 930-8555, Japan
}
\author{Osamu~Seto}
\email{seto@particle.sci.hokudai.ac.jp}
\affiliation{
  Office of International Affairs, 
  Hokkaido University, Sapporo 060-0815, Japan
}
\affiliation{
  Department of Physics, Graduate School of Science,
  Hokkaido University, Sapporo 060-0810, Japan
}
\affiliation{
Department of Life Science and Technology, 
Hokkai-Gakuen University, Sapporo 062-8605, Japan
}


\begin{abstract}

We discuss phenomenological aspects of models whose scalar
sector is extended by
an isospin doublet scalar and a complex singlet scalar
as an effective theory of supersymmetric models with mixed sneutrinos.
In such models, 
the lighter of the mixed neutral scalars
can become a viable dark matter candidate
by imposing a global $U(1)$ symmetry.
We find that the thermal WIMP scenario is 
consistent with the cosmological dark matter abundance
when the mass of the scalar is
half of that of the discovered Higgs boson
or larger than around $100~\mathrm{GeV}$.
We also point out that, 
with an additional isospin singlet Majorana fermion mediator, 
even the mass of the scalar WIMP less than around $5~\mathrm{GeV}$
is compatible with the observed dark matter abundance.
We show that such cosmologically allowed regions can be 
explored at future collider experiments and dark matter detections.

\end{abstract}
\maketitle

\cleardoublepage
\pagenumbering{arabic}

\section{Introduction}


The discovery of the Higgs boson
with a mass of $125~\mathrm{GeV}$~\cite{Aad:2012tfa,Chatrchyan:2012xdj}
and measurements of its properties
at the CERN Large Hadron Collider (LHC) 
have established the Higgs mechanism, on which
the Standard Model (SM) of particle physics is based.
Although no clear evidence against the SM has been found at collider experiments,
several phenomena that necessitate a new theory
underlying the SM have been reported mainly from the cosmological
and astrophysical observations.
These include the existence of dark matter (DM),
the baryon asymmetry of the Universe,
the cosmic inflation as well as neutrino oscillations.
We are obliged to
build new models beyond the SM and to develop methods for distinguishing 
them in order to approach the fundamental theory.


Weakly Interacting Massive Particles (WIMPs) $\chi$ are
ones of the most promising candidates for the DM in our Universe,
whose abundance is determined to be $\Omega h^2 = 0.1$
by the data obtained at the WMAP~\cite{WMAP} and Planck 
observations~\cite{Ade:2013zuv}.
If the thermal relic abundance of the WIMPs coincides with the observed value 
above, the energy scale of the new model containing the WIMP is
set at around the terascale.
Therefore, WIMP models can be explored by using data
obtained at the operative LHC Run-II and the
future electron-positron colliders, such as International
Linear Collider (ILC)~\cite{ILC,Asner:2013psa,Moortgat-Picka:2015yla,Fujii:2015jha},
the Compact LInear Collider (CLIC)~\cite{CLIC} 
and the Future Circular Collider of electrons and positrons 
(FCC-ee)~\cite{FCC-ee},
as well as many DM direct and indirect detection experiments.


Let us take a closer look at the nature of WIMPs.
As suggested by its name,
WIMPs appear to have the weak interactions with other SM particles.
However, such WIMPs that couple with the $Z$-boson have
too large annihilation cross sections
and scattering cross sections with nuclei.
The latter property completely conflicts with the null results 
of direct DM search experiments,
hence the coupling between the WIMP and the $Z$-boson 
should be absent or strongly suppressed. 
For Majorana WIMPs with the $SU(2)$ gauge interaction,
the coupling with the $Z$-boson does not exist.
For scalar WIMPs, similarly,
introduction of a $CP$ violating term in the scalar potential removes
the coupling with the $Z$-boson.
Such a prescription is common in order to avoid 
the direct DM search constraints in the literature, 
for example, in the inert doublet scalar
model~\cite{Deshpande:1977rw,Barbieri:2006dq}.
On the contrary, $SU(2)$ singlet WIMPs, 
by definition, do not interact with the $Z$-boson. 
Thus, small scattering cross sections with nuclei are predicted, 
and the direct DM search limit has been relatively easily avoided.
As one of the simplest models,
singlet scalar DM models have been intensively investigated
as the operator between
real singlet scalar WIMP and the SM Higgs doublet $H$,
$\mathcal{O} = \lambda_{H\chi}^{} |H|^2 \chi^2$, is allowed by 
renormalizability~\cite{Burgess:2000yq,Davoudiasl:2004be,Goodman:2010ku}.
Nevertheless, due to recent great progress in direct DM search experiments,
significant constraints on the coupling $\lambda_{H\chi}^{}$ 
have been imposed even for singlet DM.
On the other hand, small $\lambda_{H\chi}^{}$ allowed by
the direct detection constraints leads to the overabundance of 
the WIMP DM if one relies on the thermal WIMP paradigm mentioned above.
From the above observations about the WIMP coupling with (without) the $Z$-boson,
$SU(2)$ doublet (singlet) WIMPs end up with under-(over-)abundance
in the Universe~\cite{Arina:2007tm}.
Therefore,
one can envisage that doublet-singlet mixed WIMPs may have
the correct relic abundance and 
be consistent with the current direct DM search results simultaneously
for CP-conserving scalar or Dirac fermion DM cases in, {\it e.g.}, Ref.~\cite{Belanger:2012vp}
\footnote{Doublet-singlet Majorana WIMP models have been investigated 
in, {\it e.g.}, Refs.~\cite{Cohen:2011ec,Banerjee:2016hsk}.
In , {\it e.g.}, Refs.~\cite{Cohen:2011ec,Kadastik:2009cu,Kadastik:2009dj,Kadastik:2009gx,Bonilla:2014xba}, 
doublet-singlet mixed real scalar WIMP models with an additional $Z_2$ symmetry also have been studied.
Mixed complex scalar WIMP scenarios we focus on are qualitatively different from such CP-violating scalar or Majorana fermion WIMP scenarios.}.
For example, a supersymmetric (SUSY) model in which left-handed
and right-handed sneutrinos significantly mix due to the large 
sneutrino trilinear coupling falls in this 
category~\cite{ArkaniHamed:2000bq,Arina:2007tm,Belanger:2010cd,Kakizaki:2015nua,Arina:2015uea}.


In this Paper, we investigate the phenomenology
of scalar-type doublet-singlet mixed DM models
as simplified models of the mixed sneutrino DM models:
Since the quantum numbers of the introduced scalar doublet (singlet) 
in this Paper is the same as those of the left-handed sleptons
(right-handed sneutrinos),
the mixed sneutrino models are reduced to our model 
if all the other superparticles decouple and 
SUSY relations among couplings are relaxed.
Hence, while our model is strongly motivated by the mixed sneutrino models,
differences that stem from the absence of SUSY relations among couplings
can also be easily found by comparing the results of this Paper 
with those of previous studies on mixed sneutrino models. 
We examine the parameter space consistent with the cosmological
DM abundance.
The allowed regions are found when
the mass of the WIMP is half of that of the Higgs boson or
larger than around $100~\mathrm{GeV}$.
If we further introduce a Majorana mediator that corresponds to
the bino in the mixed sneutrino models,
the mass of the thermal WIMP can be lighter than around $5~\mathrm{GeV}$.
Then, we discuss phenomenological implications for future collider
experiments and DM detection experiments.


This paper is organized as follows.
We introduce our model in the Sec.~\ref{Sec:Model}.
After summarizing all constraints on the model in Sec.~\ref{Sec:Constraints},
we show its phenomenological implications in Sec.~\ref{Sec:Results}.
In Sec.~\ref{Sec:Model2}, our model is extended by a light Majorana
mediator for enhancing the WIMP annihilation.
We show phenomenological consequences also for the new viable parameter region.
Section~\ref{Sec:Conclusions} is devoted to concluding remarks.

\section{The Model}
\label{Sec:Model}

We introduce a complex scalar isospin doublet $\eta$ and singlet $s$
in addition to the SM particle contents.
These new fields, $s$ and $\eta$,
are charged under a new global $U(1)_X^{}$ symmetry.
The SM Higgs doublet $H$ is neutral under the $U(1)_X^{}$ symmetry.
The quantum numbers of the electroweak fields above are listed in 
Table~\ref{tab:fields}.
The $U(1)_X^{}$ charge corresponds to the dark matter number, 
and  the $U(1)_X^{}$ symmetry guarantees the stability of the lightest additional particle.
Then, the scalar potential allowed by the 
$SU(3)_C^{} \otimes SU(2)_L^{} \otimes U(1)_Y^{} \otimes U(1)_X^{}$ symmetries
and renormalizability is written as 
\begin{eqnarray}
  V&=& \mu_H^2 (H^{\dagger} H)  + \frac{\lambda_1}{2} (H^{\dagger} H)^2
       + \mu_{\eta}^2 (\eta^{\dagger} \eta) 
       +\frac{ \lambda_2}{2} (\eta^{\dagger} \eta)^2 + \lambda_3(H^{\dagger}H) (\eta^{\dagger}\eta)
       + \lambda_4(H^{\dagger} \eta)(\eta^{\dagger}H)
       \nonumber \\
   &&+ \mu_s^2 (s^* s) + \frac{\lambda_s}{2} (s^* s)^2  +\lambda_{H s}(H^{\dagger}H) (s^*s)
      + \lambda_{\eta s}  (\eta^{\dagger} \eta) (s^* s) 
      + A (\eta^{\dagger} H s + {\rm h.c.}) ,
\end{eqnarray}
where $\mu^2$'s, $\lambda$'s and $A$
are mass and coupling parameters.
It should be noticed that, because of the $U(1)_X^{}$ symmetry,
the scalar potential does not contain the $CP$ violating operator,
$\mathcal{O} = \lambda_5^{} (H^{\dagger} \eta )^2 + \mathrm{h.c.}$,
which is usually considered in the two Higgs doublet models.

\begin{table}
  \begin{center}
    \caption{The quantum numbers of the electroweak fields 
      in the mixed complex scalar WIMP model.}
    \label{tab:fields}
    \begin{tabular}{|l||c|c|c||c|}
      \hline
      Fields &$ SU(3)_C^{}$ & $SU(2)_L^{}$ & $U(1)_Y^{}$  & $U(1)_X$ \\ \hline \hline
      Left-handed lepton doublets ($L_i^{}$)  &{\bf 1}    & {\bf 2}    & $-1/2$    & $0$ \\ \hline
      Right-handed lepton singlets ($e_i^{}$) &{\bf 1}    & {\bf 1}    & $-1$ 	   & $0$ \\ \hline
      SM Higgs doublet ($H$)         &{\bf 1}    & {\bf 2}    & $+1/2$    & $0$  \\ \hline\hline
      Inert scalar doublet ($\eta$)  &{\bf 1}    & {\bf 2}    & $+1/2$    & $+1$ \\ \hline
      Inert scalar singlet ($s$)     &{\bf 1}    & {\bf 1}    & $0$ 	   & $+1$ \\ \hline
    \end{tabular}
  \end{center}
\end{table}
 
After the electroweak symmetry breaking,
the neutral component of the SM Higgs doublet develops
a vacuum expectation value, $\langle H^0 \rangle = v/\sqrt{2}$,
with $v= 246\ \mathrm{GeV}$.
Then, the mass squared matrix of the neutral component of 
the inert doublet $\eta^0$ and the inert singlet $s$ 
in the $(\eta^0, s)$ basis is diagonalized as
\begin{eqnarray}
\label{eq:inert mass matrix}
  \begin{pmatrix}
    m^2_{11} & m^2_{12} \\
    m^2_{21} & m^2_{22}
  \end{pmatrix}
& \equiv &
  \begin{pmatrix}
    \mu_{\eta}^2 + \frac{v^2}{2} \lambda & \frac{v}{\sqrt{2}} A \\
    \frac{v}{\sqrt{2}} A & \mu_s^2 +\frac{v^2}{2}\lambda_{H s}^{}
  \end{pmatrix}
\nonumber \\
& = &
  \begin{pmatrix}
    \cos \theta_{\chi}^{} & - \sin \theta_{\chi}^{} \\
    \sin \theta_{\chi}^{} & \cos \theta_{\chi}^{} 
  \end{pmatrix}
  \begin{pmatrix}
    m_{\chi_2}^2 & 0 \\
    0 & m_{\chi_1}^2 
  \end{pmatrix}
  \begin{pmatrix}
    \cos \theta_{\chi}^{} & \sin \theta_{\chi}^{} \\
    - \sin \theta_{\chi}^{} & \cos \theta_{\chi}^{} 
  \end{pmatrix} ,
\end{eqnarray}
with $\lambda \equiv \lambda_3^{} +\lambda_4^{}$.
The mass eigenvalues and the mixing angle satisfy
$m_{\chi_1} < m_{\chi_2}$ and $-\pi/2 < \theta_\chi^{} < \pi/2$.
The lighter state $\chi_1^{}$ is stable and identified with 
the WIMP candidate in our model.
The masses of the charged components $\eta^{\pm}$ are given by
\begin{equation}
\label{eq:charged scalar mass}
  m_{\eta^{\pm}}^2 = \mu_{\eta}^2 + \frac{v^2}{2} \lambda_3^{}.
\end{equation}

Ones of the most important interactions of the WIMP $\chi_1^{}$ 
in our analysis are the couplings to the $Z$-boson and to the SM
Higgs boson $h$, which depend on the mixing angle $\theta_\chi^{}$ as
\begin{eqnarray}
  \mathcal{L} \supset - i \frac{e}{\sin 2 \theta_W^{}}
 \sin^2 \theta_\chi^{} (\chi_1^* \overleftrightarrow{\partial_\mu^{}}
\chi_1^{} ) Z^\mu 
  + \left( -v \lambda \sin^2 \theta_{\chi}^{}
  - v \lambda_{H s}^{} \cos^2 \theta_{\chi}^{}
  + \frac{A}{\sqrt{2}} \sin 2 \theta_{\chi}^{} \right) h \chi_1^* \chi_1^{} .
\label{eq:hchichi}
\end{eqnarray}

\section{Experimental constraints}
\label{Sec:Constraints}

Here we discuss experimental constraints on the parameter space of our 
model.
In the framework of the standard thermal WIMP production scenario,
the DM abundance as well as direct and indirect DM 
detection results impose significant constraints on the WIMP properties.
Null results of collider searches for new particles
also considerably constrain the model parameter space.
We list notable experimental bounds 
adopted in our analysis in Table~\ref{tab:exp},
and describe them below.

\begin{table}[t]
  \begin{center}
    \caption{The experimental bounds adopted in our analysis.}
    \begin{tabular}{|c||c|c|}
      \hline
      Observable & Experimental bound \\ \hline \hline
      $\Omega h^2$ & $0.1196 \pm 0.0062\ (95\%\ \mathrm{CL})$~\cite{Ade:2013zuv}\\
      \hline
      $\sigma_{\rm Nucleon}^{} $ & LUX~\cite{Akerib:2015rjg,Akerib:2016vxi}, 
                                   CDMSlite~\cite{Agnese:2015nto} \\ 
      \hline
      $\langle \sigma_{\rm ann}^{} v \rangle$ & 
      Fermi-LAT~\cite{Ackermann:2013yva,Ackermann:2015zua,Caputo:2016ryl} \\
      \hline
      $\Gamma (Z \rightarrow \mathrm{inv.} )$ & $< 2.0\ {\rm MeV} \ (95\%\ \mathrm{CL})$~\cite{ALEPH:2005ab} \\
      \hline
      $\mathrm{Br}(h \rightarrow \mathrm{inv.} )$ & $< 0.23 \ (95\%\ \mathrm{CL})$~\cite{Aad:2015pla,Chatrchyan:2014tja} \\ 
      \hline
    \end{tabular}
    \label{tab:exp}
  \end{center}
\end{table}

The DM relic density $\Omega h^2$ is determined
through cosmological observations, most notably by 
WMAP~\cite{WMAP} and Planck~\cite{Ade:2013zuv}.
Taking the possibility that nonthermal WIMP production
contributes to the WIMP abundance into account,
we consider the value of $\Omega h^2$ shown in Table~\ref{tab:exp}
as the upper limit of the thermal WIMP abundance.
The thermal WIMP abundance is controlled by WIMP annihilation cross sections.
In most of the parameter region of this inert scalar model,
dominant WIMP annihilation modes are $\chi_1\chi_1^* \rightarrow b\bar{b}$
and $\chi_1\chi_1^*\rightarrow W^+W^-$ processes.
This argument puts the lower bound to the mixing angle $\theta_\chi^{}$.
We also explore parameter regions where
coannihilation processes become important.

Direct DM detection experiments search for
signals by the recoil energy through WIMP scattering off nuclei.
With null results, 
the expected number of events by WIMPs 
in each experiment set the upper bound on
the scattering cross section of the WIMP 
with a nucleon $\sigma_{\rm{Nucleon}}^{}$.
For a WIMP with a mass of the order of $\mathcal{O}(100)~\mathrm{GeV}$, 
recent results obtained at the LUX experiment
constrain the WIMP-nucleon cross section as 
$\sigma_{\rm{Nucleon}}^{} \lesssim \mathcal{O}(10^{-46})\ \mathrm{cm^2}$~\cite{Akerib:2016vxi}.
For a WIMP with a mass around $5~\mathrm{GeV}$,
the CDMSlite experiment imposes the most stringent upper limit as
$\sigma_{\rm{Nucleon}}^{} \lesssim \mathcal{O}(10^{-41})\ \mathrm{cm^2}$~\cite{Agnese:2015nto}.
In our model, 
the spin independent cross section of $\chi_1^{}$ 
is mediated by the Higgs boson and $Z$-boson.

The measurements of fluxes of various cosmic rays serve as 
indirect DM searches.
The most stringent limit on the DM annihilation cross section
has been obtained from diffuse $\gamma$-ray flux from
dwarf spheroidal galaxies by 
Fermi-LAT~\cite{Ackermann:2013yva,Ackermann:2015zua}.
No $\gamma$-ray signal from DM annihilation
in the Small Magellanic Cloud puts 
a similar bound on the WIMP annihilation cross section~\cite{Caputo:2016ryl}.
In addition, Super-Kamiokande results can impose upper limits on
WIMP-nucleon scattering cross sections
by non-observation of neutrinos from WIMP annihilation in the Sun~\cite{Choi:2015ara},
which might be important for WIMPs with a mass less than around 
$10~\mathrm{GeV}$.
However, those indirect search constraints are not so stringent as others.
Therefore, these indirect search limits do not explicitly
appear in our later plots although we take them into account.

Let us turn to constraints obtained from collider experiments.
If the mass of $\chi_1^{}$ is smaller than half of the mass of the $Z$-boson
(the discovered Higgs boson), the $Z$-boson (the Higgs boson) 
can decay invisibly into a pair of $\chi_1$.
The invisible decays of the $Z$- and Higgs bosons have been searched for
at LEP~\cite{ALEPH:2005ab} and LHC~\cite{Aad:2015pla,Chatrchyan:2014tja}.
Since these decay widths are proportional to $\sin^4 \theta_{\chi}^{}$,
the null results impose the upper limit on $\sin \theta_{\chi}^{}$.
Moreover, collider experiments set bounds on the masses and 
couplings of yet-to-be-discovered particles though
processes associated with WIMPs.
For example, if the charged inert scalars $\eta^{\pm}$ are once produced,
they decay into $W^{\pm (*)}$ and the missing $\chi_1^{}$.
From Eqs.~(\ref{eq:inert mass matrix}) and (\ref{eq:charged scalar mass}), the squared mass difference for small mixing angle, $\theta_{\chi}^{} \ll1$, 
is given by $m_{\chi_2^{}}^2-m_{\eta}^2 \simeq v^2_{} \lambda_4^{}/2$.
In our analysis, the reference value of $\lambda_4$ is as small as 
the corresponding scalar coupling constant derived from
the SU(2) D-terms in the mixed sneutrino models~\cite{ArkaniHamed:2000bq,Arina:2007tm,Belanger:2010cd,Kakizaki:2015nua,Arina:2015uea} as
$\lambda_4 = 2m_Z^2 \cos^2\theta_W \cos^2\beta / v^2\simeq 2.3\times 10^
{-5}$
with $\tan\beta=10$.
Therefore, the charged and heavy neutral scalars are sufficiently 
degenerate
in mass, leading to negligible contributions to the $T$ parameter,
$\Delta T \sim 0$.

\section{Numerical results}
\label{Sec:Results}

With the aid of {\tt LanHEP}~\cite{LanHEP}, which automatically
generates Feynman rules,
we implement our mixed complex scalar WIMP model into 
the public codes {\tt micrOMEGAs}~\cite{Belanger:2013oya}
and {\tt CalcHEP}~\cite{Belyaev:2012qa}, 
which allows for automated computations
of the properties of DM and associated new particles.
All numerical results presented here are obtained with {\tt micrOMEGAs}
and {\tt CalcHEP}.
As benchmark scenarios in our analysis,
we take the scan bounds and reference values listed in
Table~\ref{tab:benchmark}.
It should be noticed that the $h \chi_1^{} \chi_1^*$-coupling 
given in Eq.~(\ref{eq:hchichi}) controls
both the WIMP-nucleon scattering cross section and annihilation cross sections.
Thus, the effect of 
the variation of $\lambda$ or $\lambda_{H s}^{}$ can be absorbed by the change
of the third term,
and thus the viable mass range of $\chi_1^{}$ is not altered.
Moreover, the first term in the $h \chi_1^{} \chi_1^*$-coupling
is strongly suppressed by $\sin^2 \theta_{\chi}^{}$,
and thus negligible in many cases.
From this observation, 
we fix the values of the parameters $\lambda$ and $\lambda_{H s}^{}$ 
at those motivated in the mixed sneutrino WIMP
scenarios~\cite{ArkaniHamed:2000bq,Arina:2007tm,Belanger:2010cd,Kakizaki:2015nua,Arina:2015uea}, where $\lambda = (m_Z^2/v^2) \cos 2 \beta$
and $\lambda_{H s}^{} = |y_\nu^{}|^2 \simeq 0$~{
In the small $\theta_{\chi}^{}$ limit with finite $\lambda_{H s}^{}$,
the second term of 
$h\chi_1^{}\chi_1^*$-coupling in Eq.~(\ref{eq:hchichi}) becomes 
the most relevant,
hence this model is reduced to the so-called Higgs portal 
singlet scalar DM model.}.
For $\tan \beta = 10$, we obtain $\lambda = - 0.14$.
Given the experimental constraints discussed in the previous
section, we find two classes of allowed parameter region:
(A) $m_{\chi_1^{}}^{} \simeq m_h^{}/2$ (Higgs-pole region); 
and (B) $m_{\chi_1^{}}^{} \gtrsim 100~\mathrm{GeV}$ (Large WIMP mass region).

\begin{table}[t]
  \begin{center}
    \caption{The scan bounds and reference values of parameters of our model.}
    \label{tab:benchmark}
    \begin{tabular}{|c||c|}
      \hline
      Parameter		&	Scan bound / Reference value 	\\ \hline \hline
      $m_{\chi_1}^{}$		& [$10$ MeV, $1.5$ TeV]\\ \hline
      $m_{\chi_2}^{}$	& [$100~\mathrm{GeV}$, $16.5$ TeV]	\\ \hline
      $\sin\theta_{\chi}^{}$	& [$0.001$, $1$]\\ \hline
      $m_{\psi}^{}$ 	& [$10$ MeV, $1.01~\mathrm{TeV}$]	\\ \hline \hline
      $\lambda$	& $- 0.14$ \\ \hline
      $\lambda_{H s}^{}$	& $0$\\ \hline
    \end{tabular}
  \end{center}
\end{table}

\subsection{Higgs-pole region} 

\begin{figure}[t]
  \begin{center}
    \includegraphics[clip,width=14cm]{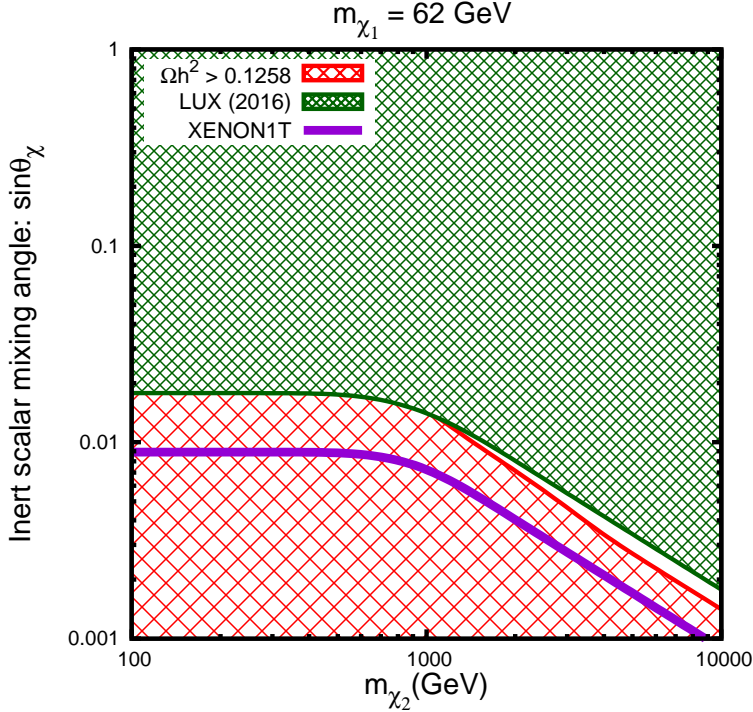}
    \caption{Experimental constraints and future prospects 
      in the $(m_{\chi_2^{}}^{}, \sin \theta_\chi^{})$
      plane 
      in the complex scalar for $m_{\chi_1^{}}^{} = 62~\mathrm{GeV}$.
    }
    \label{fig:higgspole}
  \end{center}
\end{figure}

A viable parameter region can be found
when the WIMP annihilation in the early Universe
takes place near the Higgs pole, 
namely $m_{\chi_1^{}}^{}\simeq m_h^{}/2$.
Figure~\ref{fig:higgspole} shows experimental constrains
and future prospects in the $(m_{\chi_2^{}}^{}, \sin \theta_\chi^{})$
plane for $m_{\chi_1^{}}^{} = 62~\mathrm{GeV}$.
Excluded parameter regions in the light of 
current experimental bounds are shown with mesh areas.
In the red mesh region, the resultant relic abundance exceeds
the observed DM density~\cite{Ade:2013zuv}.
The green mesh region is 
excluded by the LUX experiment~\cite{Akerib:2016vxi} .
The purple solid line indicates the expected reach by
the XENON-1T experiment~\cite{Aprile:2015uzo}.

The WIMP-nucleon scattering is induced mainly by 
the $Z$-boson exchange diagram for
for $m_{\chi_2}^{} \lesssim 1000~\mathrm{GeV}$, 
and by the Higgs boson one for
$m_{\chi_2}^{} \gtrsim 1000~\mathrm{GeV}$.
This fact leads to the break of the direct detection limit around
$m_{\chi_2}^{} \simeq 1000~\mathrm{GeV}$ in Fig.~\ref{fig:higgspole}.
Since $m_{\chi_1} \simeq m_h /2$,
the $s$-channel Higgs boson exchange process is the dominant annihilation mode
and controls the relic density.
For this process, the most relevant interaction is
the third term of $h\chi_1^{}\chi_1^*$-coupling in Eq.~(\ref{eq:hchichi}), 
which is rewritten as
\begin{equation}
  \frac{A}{\sqrt{2}}\sin 2\theta_{\chi} 
  \simeq \frac{\sin^2 2\theta_{\chi}(m_{\chi_2}^2-m_{\chi_1}^2)}{2v}.
\end{equation}
Namely, a larger mass deference between $\chi_1$ and $\chi_2$ leads to
a smaller relic density, resulting in a viable region for 
$m_{\chi_2} \gtrsim 1000~\mathrm{GeV}$,
As can be seen from Fig.~\ref{fig:higgspole},
the current allowed region can be ruled out by the XENON-1T experiment.

We comment on the vacuum stability bound.
Since the trilinear coupling $A$ is large in the allowed Higgs-pole region,
there may appear a deeper vacuum than the electroweak one.
In Ref.~\cite{Kakizaki:2015nua}, vacuum meta-stability has been investigated 
in mixed sneutrino WIMP scenarios.
It has been shown that the upper bound on the mixing angle is $\sin\theta_{\tilde{\nu}}^{} \lesssim 0.26$ 
for a WIMP mass of $m_{\tilde{\nu}}^{} \sim 1~\mathrm{GeV}$.
Since in our model the mixing angle in the allowed Higgs-pole region is as small as $\sin\theta_{\chi} \lesssim 0.01$,
we expect that the electroweak vacuum is stable enough.
Further discussion on the vacuum stability is beyond the scope of this paper.

\subsection{Large WIMP mass region} 

\begin{figure}[!t]
  \begin{center}
    \includegraphics[clip,width=14cm]{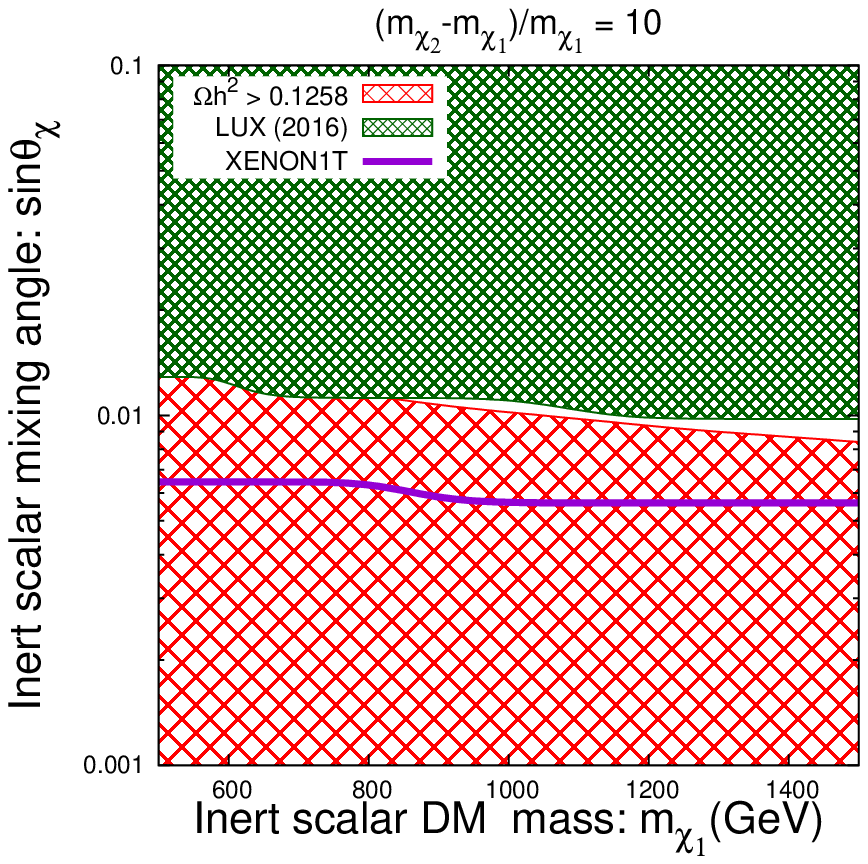}
    \includegraphics[clip,width=14cm]{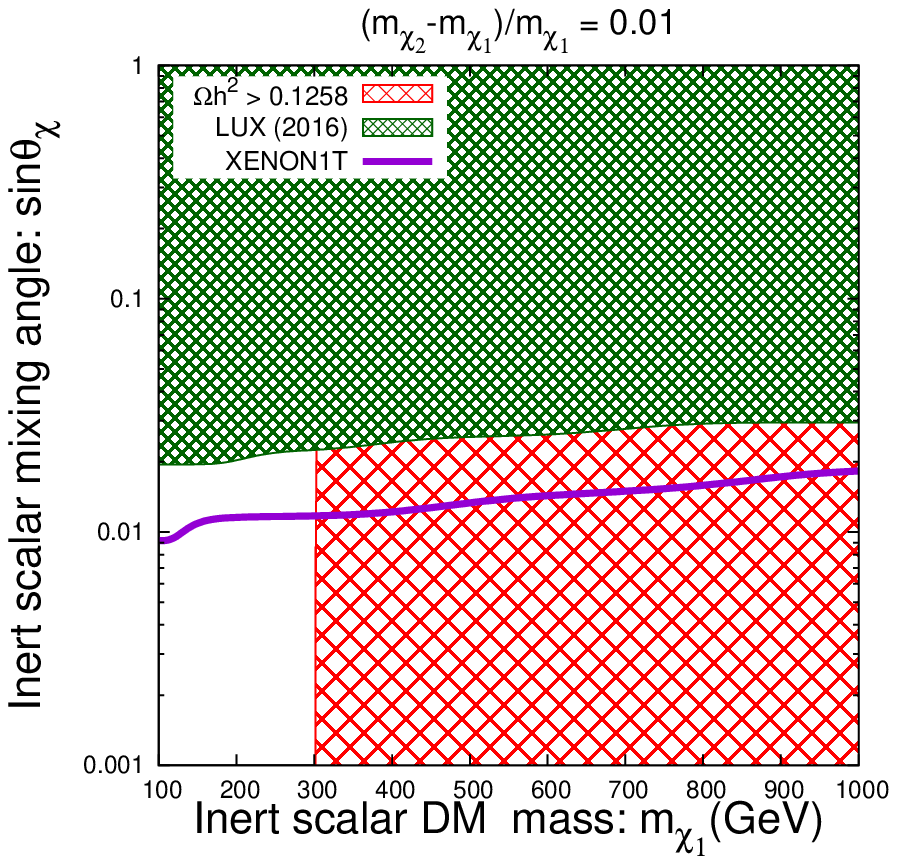}
    \caption{Experimental constraints and future prospects
      in the $(m_{\chi_1^{}}^{}, \sin \theta_\chi^{})$ plane
      in the mixed complex scalar WIMP model.
      The top (bottom) frame shows the split (degenerate) case
      for $(m_{\chi_2^{}}^{}-m_{\chi_1^{}}^{})/m_{\chi_1^{}}^{}=10$
      ($(m_{\chi_2^{}}^{}-m_{\chi_1^{}}^{})/m_{\chi_1^{}}^{}=0.01$).
    }
    \label{fig:heavywimp}
  \end{center}
\end{figure}

Constraints from direct DM searches become weaker
as the WIMP mass is increased for $m_{\chi_1^{}}^{} \gtrsim 100~\mathrm{GeV}$,
leading to another viable mass region.
Figure~\ref{fig:heavywimp} shows allowed regions
in such large WIMP mass cases
in the $(m_{\chi_1^{}}^{}, \sin \theta_\chi^{})$ plane.
There are two representative cases for decreasing the WIMP relic abundance
for $m_{\chi_1^{}}^{} \gtrsim 100~\mathrm{GeV}$:
The masses of $\chi_1^{}$ and $\chi_2^{}$ are considerably split, 
or degenerate enough to coannihilate.
In the split case,
the mass difference between $\chi_1^{}$ and $\chi_2^{}$ should be
large so that the WIMP coupling to the Higgs boson 
and the annihilation cross section are sufficiently enhanced.
In our numerical analysis, we take 
$(m_{\chi_2^{}}^{}-m_{\chi_1^{}}^{})/m_{\chi_1^{}}^{}=10$ for the split case,
and $(m_{\chi_2^{}}^{}-m_{\chi_1^{}}^{})/m_{\chi_1^{}}^{}=0.01$ for 
the degenerate case.
In this degenerate case, the charged scalar decays into $\chi_1^{}$
and too soft jets or leptons to be detected.
Therefore, the charged scalar mass $m_{\eta^{\pm}}$ is not constrained at the LHC.
The dominant coannihilation modes of
$\eta^+ \eta^- \rightarrow W^+ W^-$,
$\chi_2^{} \chi_2^{} \rightarrow W^+ W^- (Z Z)$ and
$\chi_2^{} \eta^\pm \rightarrow \gamma W^\pm$
significantly decrease the WIMP relic abundance.
These modes contribute to around $30\%$ of the effective 
annihilation cross section.
As one can see from Fig.~\ref{fig:heavywimp},
the viable parameter region in the split case (top frame) 
lies in the reach of the future XENON-1T experiment.
On the other hand, that in the degenerate case (bottom) 
is only partially covered by the expected XENON-1T sensitivity reach.
Therefore, these two cases are distinguishable 
through future direct detection experiments. 

\section{The model with a Majorana fermion}
\label{Sec:Model2}

In the previous section, 
we have shown that there are two regions consistent with current 
experimental results in our mixed complex scalar WIMP model;
the Higgs-pole region where $m_{\chi_1^{}}^{} \simeq m_h^{}/2$
and large WIMP mass region
where $m_{\chi_1^{}}^{} \gtrsim \mathcal{O} (10^2)~\mathrm{GeV}$.
Thus, a mass of around $60~\mathrm{GeV}$
appears to be the smallest for viable mixed complex scalar WIMPs.
However, in fact, a minor extension of the model
opens another viable mass range.
In this section, we show that a GeV-mass WIMP 
is also feasible by introducing only
one light Majorana fermion $\psi$ that mediates WIMP annihilation
into the model described in Sec.~\ref{Sec:Model}.

We assume that the newly introduced Majorana fermion
is totally singlet under all the SM gauge symmetry 
as well as the global $U(1)_X^{}$ symmetry.
The quantum numbers of the electroweak particles of this new model
is summarized in Table~\ref{tab:fields2}.
The left- and right-handed leptons, $L_i^{}$ and $e_i^{}$,
 have negative $U(1)_X$ charges in this model.
 Here, $i\ (=1-3)$ denotes the generation index.
Then, interaction terms among 
the inert doublet, the left-handed lepton doublets and the Majorana fermion 
are allowed with their couplings dependent on the lepton generations.
Assuming a hierarchical relation in these couplings,
we introduce only the interaction with the third generation left-handed lepton
as

\begin{eqnarray}
  \Delta \mathcal{L} = -Y \left( \overline{L}_3^{} \psi \tilde{\eta} 
  + \mathrm{h.c.} \right),
  \label{eq:yukawa}
\end{eqnarray}
with $Y$ being a Yukawa coupling constant, and $\tilde{\eta}=i\sigma_2^{} \eta^*$, where $\sigma_2^{}$ is the second Pauli matrix.
In the mass eigenstates, this interaction is expressed as
\begin{eqnarray}
  \Delta \mathcal{L} =
  \frac{Y}{2} \sin \theta_{\chi}^{} 
  [ \bar{\nu}_{\tau}(1-\gamma_5^{}) \psi \chi_1^* + \mathrm{h.c.} ] +
  \cdots.
\end{eqnarray}
The bottom line is that
with this additional interaction,
GeV-mass WIMPs can annihilate into a pair of anti-tau neutrinos through
the $t$-channel $\psi$ exchange, 
$\chi_1^{} \chi_1^{} \rightarrow \bar{\nu}_{\tau}^{}\bar{\nu}_{\tau}^{}$.
The annihilation modes into fermion and anti-fermion pairs, 
$\chi_1^{}\chi_1^* \rightarrow f\bar{f}$, contribute to the effective annihilation cross section less than $1\%$.
Since $\psi$ is an isospin singlet, the invisible decay widths of the $Z$- and Higgs bosons
to a pair of $\psi$ are absent due to the $SU(2)$ invariance.

\begin{table}
  \begin{center}
    \caption{The quantum numbers of the electroweak particles
      in the mixed complex scalar WIMP model with a Majorana fermion.}
    \label{tab:fields2}
    \begin{tabular}{|l||c|c|c||c|}
      \hline
      & $ SU(3)_C$ & $SU(2)_L$ & $U(1)_Y$  & $U(1)_X$ \\ \hline \hline
 Left-handed lepton ($L_i^{}$) &{\bf 1}    & {\bf 2}    & $-1/2$    & $-1$ \\ \hline
 Right-handed lepton ($e_i^{}$) &{\bf 1}    & {\bf 1}    & $-1$ 	   & $-1$ \\ \hline
 SM Higgs doublet ($H$)        &{\bf 1}    & {\bf 2}    & $+1/2$    & $0$  \\ \hline\hline
 Inert scalar doublet ($\eta$) &{\bf 1}    & {\bf 2}    & $+1/2$    & $+1$ \\ \hline
 Inert scalar singlet ($s$)   &{\bf 1}    & {\bf 1}    & $0$ 	   & $+1$ \\ \hline
 Majorana fermion ($\psi$)   &{\bf 1}    & {\bf 1}    & $0$ 	   & $0$  \\ \hline
    \end{tabular}
  \end{center}
\end{table}

\subsection{Experimental constraints}

The experimental constrains described in Sec.~\ref{Sec:Constraints}
also apply to the new WIMP model with a Majorana mediator.
In addition, 
the existence of the new decay modes $\eta^\pm \to \tau^\pm \psi$, which
are induced by the Yukawa interaction in Eq.(\ref{eq:yukawa}),
imposes another constraint on $\eta^\pm$.
It should be noticed that similar processes are analyzed in the context
of SUSY models. 
Non-observation of signals of two leptons plus a missing energy
sets bounds on slepton 
masses~\cite{LEPresults,Aad:2014yka,Aad:2014vma,Khachatryan:2014qwa}.
Searches for the stau $\tilde{\tau}$ through the decay 
into the lightest neutralino $\tilde{\chi}^0$,  
$\tilde{\tau} \rightarrow \tau \tilde{\chi}^0_1$, 
at the LHC experiment impose the lower limit on the mass of the stau 
as~\cite{Aad:2014yka}
\begin{equation}
  m_{\tilde{\tau}_L^{}}^{} >93.1~\mathrm{GeV} \quad (95\%\ \mathrm{CL}) ,
\end{equation}
for a massless neutralino.
For simplicity, we apply this bound on the mass of 
the charged inert scalar $\eta^{\pm}$ in our model.

\subsection{Numerical results}

\begin{figure}[t]
  \begin{center}
    \includegraphics[clip,width=14cm]{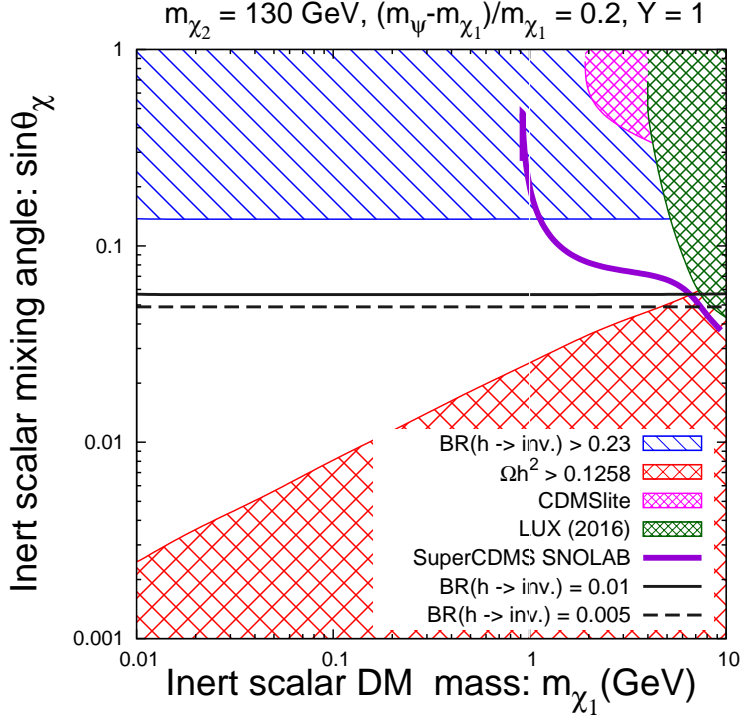}
    \caption{Experimental constraints and future prospects
      in the $(m_{\chi_1^{}}^{},\sin \theta_\chi^{})$ plane
      in the model with a Majorana fermion for 
      $m_{\chi_2^{}}^{} =130~\mathrm{GeV}$,
      $(m_\psi^{}-m_{\chi_1^{}}^{})/m_{\chi_1^{}}^{}=0.2$ and $Y=1$.}
    \label{fig:gev-mass}
  \end{center}
\end{figure}

\begin{figure}[t]
  \begin{center}
    \includegraphics[clip,width=14cm]{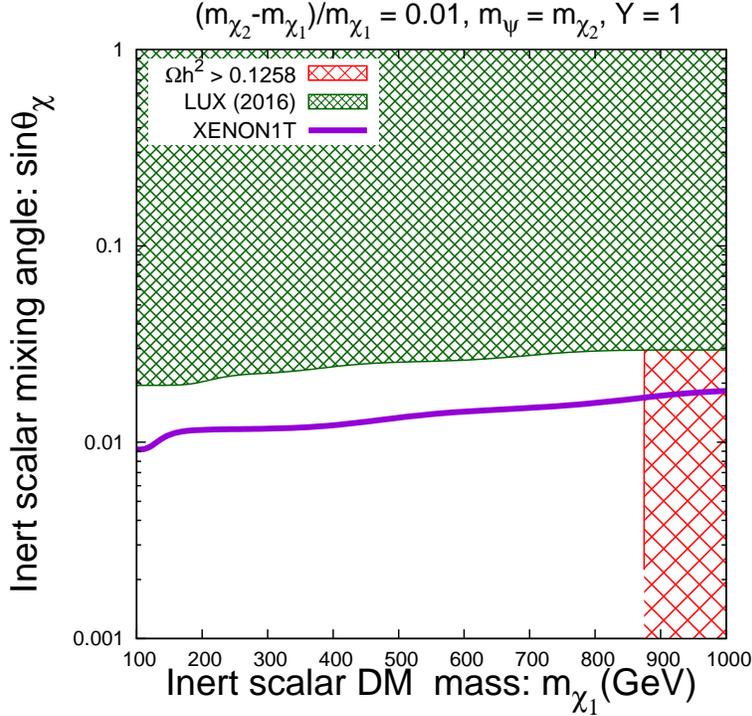}
    \caption{Experimental constraints and future prospects
      in the $(m_{\chi_2^{}}^{},\sin \theta_\chi^{})$ plane
      in the model with a Majorana fermion
      for $(m_{\chi_2^{}}^{}-m_{\chi_1^{}}^{})/m_{\chi_1^{}}^{}=0.01$,
      $m_\psi^{} = m_{\chi_2}^{}$ and $Y=1$.
    }
    \label{fig:heavydegwimp}
  \end{center}
\end{figure}

Our numerical results in the model with the Majorana mediator are
shown in Fig.~\ref{fig:gev-mass} for $m_{\chi_1^{}}^{} < 10~\mathrm{GeV}$ 
and in Fig.~\ref{fig:heavydegwimp} for 
$100~\mathrm{GeV} < m_{\chi_1^{}}^{} < 1000~\mathrm{GeV}$.

In Fig.~\ref{fig:gev-mass}, experimental constraints
and future prospects are shown in the 
$(m_{\chi_1^{}}^{}, \sin \theta_{\chi}^{})$ plane
for $m_{\chi_2}^{} = 130~\mathrm{GeV}$, 
$(m_\psi^{}-m_{\chi_1^{}}^{})/m_{\chi_1^{}}^{}=0.2$ and $Y=1$.
The blue hatched region is excluded by the null results of
the invisible decay of the Higgs boson 
at the LHC~\cite{Aad:2015pla,Chatrchyan:2014tja}.
The region where the thermal WIMP abundance is overabundant
is covered with red mesh~\cite{Ade:2013zuv}.
The excluded regions by the CDMSlite experiment~\cite{Agnese:2015nto}
and LUX experiments~\cite{Akerib:2015rjg} are covered with
magenta mesh and green mesh, respectively.
The solid purple line shows the future expected sensitivity of 
SuperCDMS SNOLAB~\cite{Cushman:2013zza}.
The solid (dashed) black line 
represents the case where the invisible decay rate of the Higgs boson
is $\mathrm{Br} (h\rightarrow \mathrm{inv.}) = 0.01~(0.005)$.
These values should be compared with the future expected sensitivity
at the ILC.
The ILC stage with $\sqrt{s}=250~\mathrm{GeV}$ and 
$L=250~\mathrm{fb^{-1}}$ aims the level of
$\mathrm{Br} (h\rightarrow \mathrm{inv.}) = 0.0069$
using the polarization configuration of 
$(P_{e^-}^{},P_{e^+}^{}) = (+80\%,-30\%)$~\cite{Ishikawa}.
In our numerical calculations, 
we take $m_{\chi_2}=130~\mathrm{GeV}$ so that
the Higgs boson does not decay into states containing invisible $\chi_2^{}$.
WIMP annihilation proceeds not only by the Higgs boson or the $Z$-boson 
but also by the $t$-channel exchange of $\psi$,
and the resultant relic abundance depends on the mass
and the coupling of the Majorana mediator.
We find that 
the effective annihilation cross section (the relic abundance) 
is maximized (minimized) for
$(m_\psi^{}-m_{\chi_1^{}}^{})/m_{\chi_1^{}}^{}=0.2$.
For smaller mass differences, coannihilation effects rather increase 
the relic abundance.
As for the Yukawa coupling, we set $Y=1$ as a reference.
Larger $Y$ relaxes the constraint from the DM relic abundance,
and vice versa.
For $m_{\chi_1^{}}^{}< 5~\mathrm{GeV}$, the LHC imposes
the strongest limit on the mixing angle as $\sin\theta_{\chi} < 0.14$.
This figure shows that the allowed region 
can be further explored by 
future DM direct detection 
and precise measurements of the invisible decay of the Higgs boson 
at future electron-positron colliders.

The introduction of the Majorana mediator also affects 
the allowed region for large WIMP mass cases 
as shown in Fig.~\ref{fig:heavydegwimp}.
We take $(m_{\chi_2^{}}^{}-m_{\chi_1^{}}^{})/m_{\chi_1^{}}^{}=0.01$, 
$m_\psi^{} = m_{\chi_2}^{}$ and $Y=1$.
The extension of the allowed region compared with Fig.~\ref{fig:heavywimp}
is caused by coannihilation processes mediated by $\psi$.
The dominant coannihilation modes in this case are 
$\eta^\pm \eta^\pm \to \tau^\pm \tau^\pm$,
$\chi_2^{(*)} \eta^\pm \to \bar{\nu} \tau^+ (\nu \tau^-)$ and
$\chi_2^{(*)} \chi_2^{(*)} \to \nu \nu (\bar{\nu} \bar{\nu})$, 
whose relative contributions to 
the effective annihilation cross section amount to around $75\%$.
This allowed region is further investigated at XENON-1T~\cite{Aprile:2015uzo}.

\section{Conclusions}
\label{Sec:Conclusions}

We have investigated phenomenological implications of the complex
scalar WIMP that is an admixture of an
isospin doublet scalar and a complex singlet scalar.
This class of model is naturally realized 
in SUSY models equipped with right-handed sneutrinos that
have large trilinear scalar couplings.
Due to a hypothetical global $U(1)$ symmetry,
the lighter mixed neutral scalar is stabilized,
and thus become a WIMP.
We have shown that there are two viable WIMP mass ranges
where the WIMP abundance is consistent with the DM abundance:
$m_{\chi_1^{}}^{} \simeq m_h/2$ and
$m_{\chi_1^{}}^{} \gtrsim 100~\mathrm{GeV}$.
We have also pointed out that 
by introducing an additional isospin singlet Majorana fermion,
the constraint from the dark matter abundance can be satisfied
even when the mass of the WIMP is smaller than around $5~\mathrm{GeV}$.
These cosmologically allowed regions can be 
further probed at upgraded DM detection experiments
and future collider experiments.

\begin{acknowledgments}
We thank Shinya Kanemura, Hiroaki Sugiyama and Koji Tsumura 
for valuable discussions and comments.
This work was supported, in part, by 
Grant-in-Aid for Scientific Research on Innovative Areas,
the Ministry of Education, Culture, Sports, Science and Technology,
Nos.\ 16H01093 (M.K.) and 26105514 (O.S.),
Grant-in-Aid for Scientific Research (C),
Japan Society for the Promotion of Science, No.\ 26400243 (O.S.), 
the Sasakawa Scientific Research Grant from The Japan Science Society (A.S.),
and by the SUHARA Memorial Foundation (O.S.).
\end{acknowledgments}

\end{document}